\documentclass[reprint,preprintnumbers, amsmath,amssymb, aps]{revtex4-2}

\usepackage{graphicx}
\usepackage{dcolumn}
\usepackage{bm}
\usepackage{changes}
\usepackage{soul}
\usepackage{amsmath}
\usepackage{color}

\begin{document}

\title{Continuous coherent perfect absorption and lasing at an exceptional point of anti-parity-time symmetric photonic structures}

\author{Jeng Yi Lee}

\affiliation{Department of Opto-Electronic Engineering, National Dong Hwa University, Hualien 974301, Taiwan}

\begin{abstract}
 We consider a type of hypothetical compound materials  in which its refractive index in spatial distribution meet $n(-x)=-n^{*}(x)$, belonging to anti-parity-time (APT) symmetric structures.
 Additionally, we demand balanced real positive- and negative- permeabilities with $\mu(-x)=-\mu(x)$.
 By introducing parametrization into APT symmetric transfer matrix, together with reciprocity theorem, we propose a generic parametric space to display its associated scattering results including symmetry phase, exceptional point, and  symmetry broken phase.
 The outcome is irrespective of any system complexity, geometries, materials, and operating frequency.
With the parametric space, we find that APT symmetric system not only enables coherent perfect absorption or lasing occurred at an exceptional point, but also realize a simultaneous coherent perfect absorption-lasing.
Since APT-symmetric system is constructed by balanced positive and negative index materials, the phase accumulated from optical path length is null, resulting in an assignment of mode order lost. 
To verify our analysis, several designed heterostructures are demonstrated to support our findings.
\end{abstract}

\maketitle

\textit{Introduction-} Negative index materials (NIMs) whose both permittivity and permeability have real negative  values can exhibit unusual electromagnetic responses as well as provide a fertile ground for designs of exotic functional devices, such as subwavelength near-field imaging \cite{superlens}, invisible cloaks \cite{invisible}, to name a few.

On the other hand, non-Hermitian of open systems involving material loss, material gain, or both, violated by the unitary relation for scattering matrix,  has recently attracted attention due to its unusual wave manipulation \cite{nonher1,nonher2,nonher3}.
One achievement in this witness is parity-time (PT) symmetric systems with balanced gain and loss, whose
its scattering matrix of $S$ obeys $PTSPT=S^{-1}$ valid for any complexity of system configuration and its dimension \cite{yidong,nonher3}.
Here $PT$ is a combined parity-time symmetry operation.
Such scattering matrix relation leads to different responses in scattering eigenvalues: ones are distinction and unimodulus with no net amplification and no net dissipation, corresponding to symmetry phase,  and others are in pairs of reciprocal moduli with one amplification and another dissipation, corresponding to symmetry broken phase. 
In a spontaneous phase transition, it corresponds to an exceptional point (EP) with two or more eigenvalues and its associated eigenstates coalesced.
With these unusual scattering properties, PT-symmetry systems can support coherent perfect absorption-lasing \cite{CPAL}, anisotropic transmission resonances \citep{transmission}, double refraction \cite{dref}, one-way optical pull force \cite{force1,force2}, unidirectional polarization beam splitters \cite{beamsplitter}, negative refraction \cite{negativeref}, topological edge state \cite{topology1,topology2} to name a few. 

Another achievement closely related to  aforementioned properties is anti-parity-time (APT) symmetry whose its refractive index in spatial placement satisfies $n(x)=-n^{*}(-x)$ \cite{APT1}, that is antisymmetry with $\{PT,H\}=0$ for Hamiltonian $H$.
Consequently, the real part of $n(x)$ has odd symmetry, while the imaginary part of that has even symmetry.
The latter condition indicates that APT-symmetric systems can employ gain alone, loss alone, lossless alone, or hybrid.
APT-symmetric systems enable a flat broadband light transmission \cite{APT1},  continuous lasing or coherent perfect absorption spectra \citep{APT1,APT2}, large spatial Hoos-H\"anchen sift \cite{APT4}, chiral mode conversion \cite{APT5}, chiral polarizer\cite{APT6}, coherent switch \cite{APT7},  distinct entanglement dynamics in different scattering phases  \cite{APT8}, and coherent perfect absorber-laser circuit \cite{chen1}.

\begin{figure*}[t]
\centering
\includegraphics[width=1\textwidth]{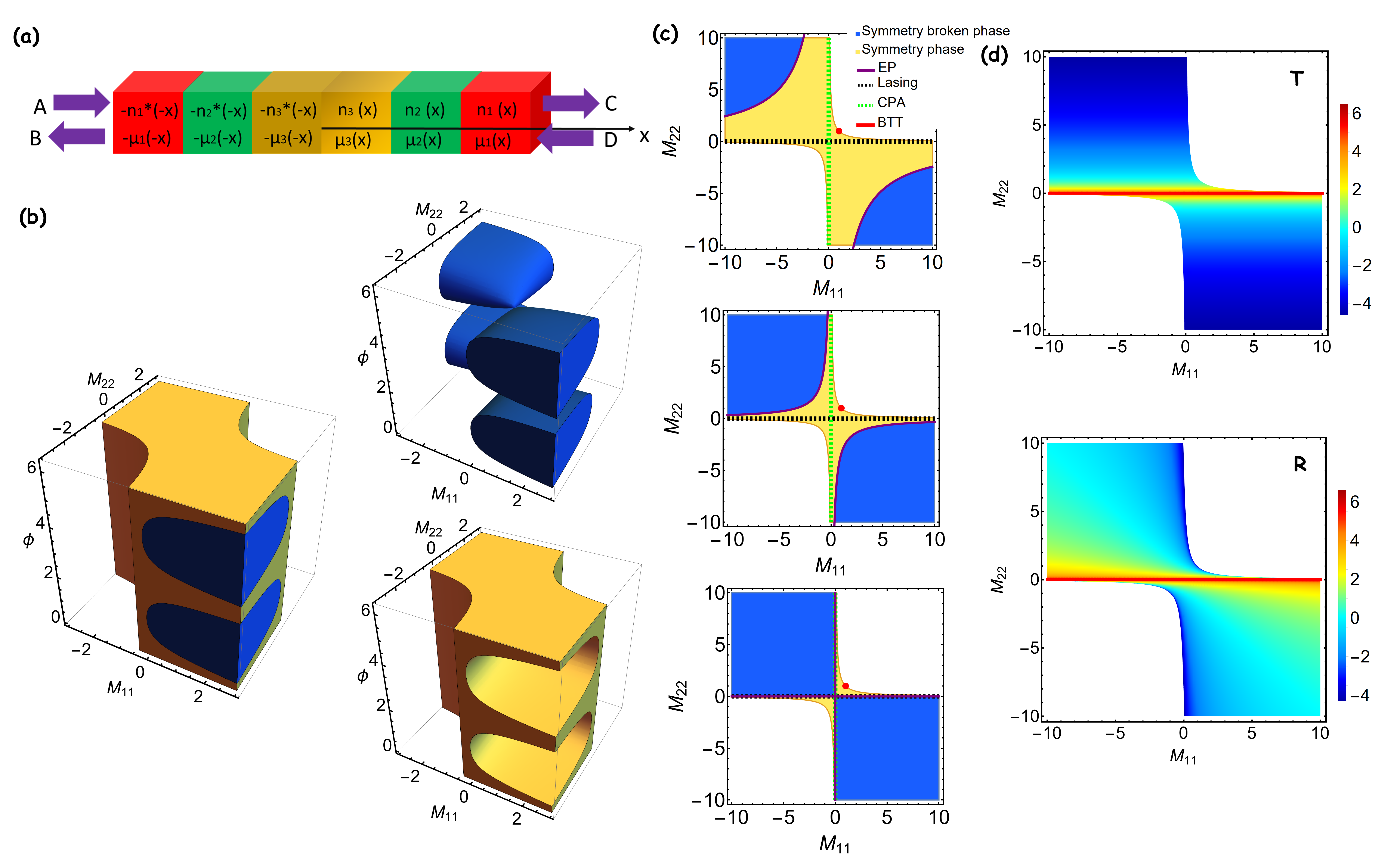}
 \caption{(a) Schematic of a APT-symmetric heterostructure having $n(x)=-n^{*}(-x)$ and $\mu(x)=-\mu(-x)$. (b) 3D parametric space for any APT-symmetric systems. We specifically mark symmetry broken phase by blue color and mark symmetry phase by yellow color. (c) With $\phi=0.2,0.5,0.5\pi$, we provide the corresponding parametric space in terms of $M_{11}$ and $M_{22}$. Between symmetry phase and broken symmetry phase, it is EP marked by a purple line. We also mark CPA, Lasing, and BTT by a dashed green line, a dashed black, and a red dot, respectively. With the parametric space, we calculate the logarithm of transmittance $log(T)$ and the logarithm of reflectance $log(R)$ in (d). Here the red lines correspond to infinite values.}
 \end{figure*}

However, it is desired to recognize comprehensive scattering properties that can benefit novel designs of functional devices.
In this work, we consider a one-dimensional (1D) APT-symmetric system with $n(x)=-n^{*}(-x)$.
We additionally require balanced real positive and negative permeabilities with $\mu(-x)=-\mu(x)$.
Since the corresponding transfer matrix  $M$ obeys $\sigma M=[M^{*}]^{-1}\sigma$, we thus introduce parametrization for $M$. Here $\sigma$ is parity operator.
Together with optical reciprocity, we propose a parametric space to display any scattering properties including not only transmittance and reflectances but also symmetry phase, exception point (EP), and symmetry broken phase. 
We observe that APT-symmetric systems allow not only  coherent perfect absorption (CPA) or lasing occurred at EP, but also coherent perfect absorption-lasing (CPAL) .
  On the other hand, with a leverage of balanced positive index materials (PIMs) and NIMs, the phase accumulated from optical path length is lost, resulting in a real transmission coefficient and an assignment of mode order lost.
We provide several designed systems to verify  our finding.

\begin{figure*}[t]
\centering
\includegraphics[width=1\textwidth]{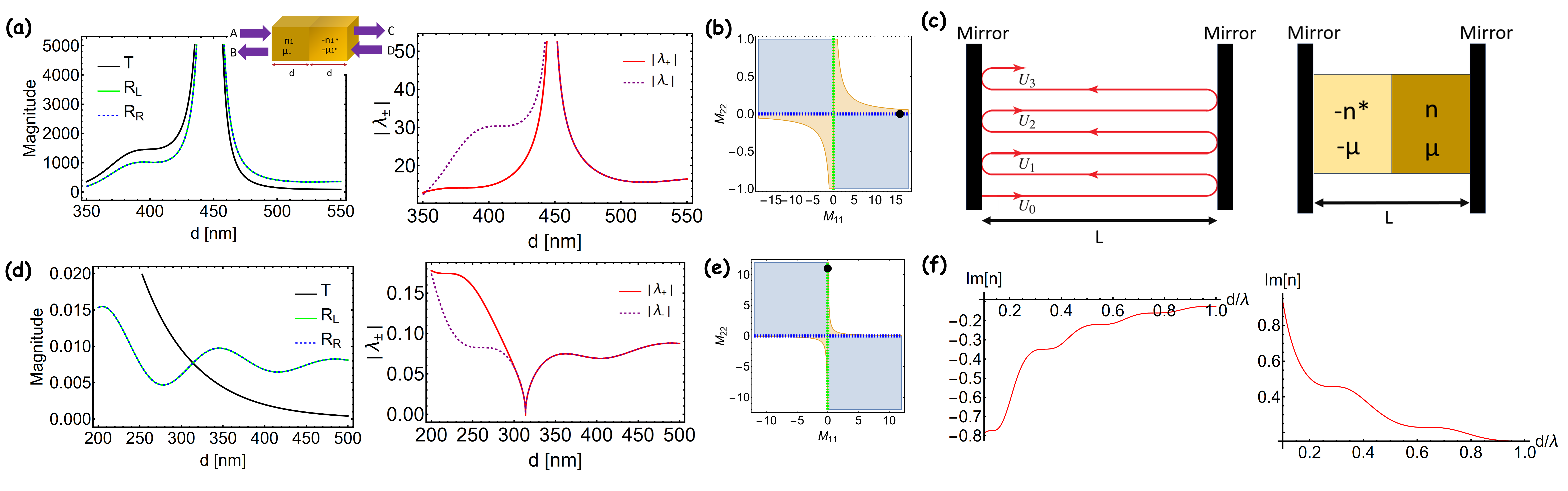}
 \caption{ (a) Transmittance, two reflectances, and two scattering eigenvalues  of bi-slab heterostructure with respect to slab length $d$. Schematic of the system configuration is shown in insert. This  heterostructure exhibits lasing at $d=450 [nm]$.  With the parametric space, we analyze the system marked by a black dot in (b).  A gain (APT-symmetric) material is encompassed by two identical lossless mirrors in left (right) of (c). Another simple bi-slab heterostructure with lossy materials embedded is designed to exhibit CPA at $d=313.85 [nm]$ in (d). With respect to $d$, the corresponding   Transmittance, two reflectances, and two eigenvalues are calculated in (d). We analyze this system with the parametric space in (e).  With fixed $Re[n]=-2.3$ ($Re[n]=-1.5$), $\lambda=450 [nm]$, and $\mu=-2$ ($\mu=-1.5$), we calculate $Im[n]$ with respect to $d$ to maintain lasing (CPA) in the left (right) of (f), exhibiting a continuous spectrum. }
 \end{figure*}

\textit{Theory-}
We consider a 1D APT-symmetric heterostructure as shown in Fig. 1 (a).
We let electromagnetic harmonic plane waves normally upon the  heterostructure with their complex amplitudes being denoted as $A$ and $D$ in the left and right leads, while the complex amplitudes for associated scattering (out-going) waves in the left and right leads are denoted as $B$ and $C$. 
APT-symmetric structures need to embed balanced PIMs and NIMs, while the imaginary part of $n(x)$ can be negative (gain) alone, positive (loss) alone, zero (lossless) alone, or hybrid.
Under operating spatial exchange of $n\leftrightarrow -n^{*}$ and $\mu\leftrightarrow  -\mu$  with respect to $x=0$, the transfer matrix $M$ becomes $M^{*}$.
It leads to 
$
\sigma M=[M^{-1}]^{*}\sigma
$ \cite{APT1}.
Here the transfer matrix $M$ is defined by $[C,D]^T=M[A,B]^T$ and $\sigma$ is parity operation.
With this relation, we apply parametrization to $M$ only with
 three  real parameters: $M_{11}$, $M_{22}$, and $\phi$.
Moreover, due to same background in the left and right leads, optical reciprocity can be applied.
As a result, we state
\begin{equation}
\begin{split} M=
\begin{bmatrix}
M_{11} & \sqrt{1-M_{11}M_{22}}e^{i\phi}\\
-\sqrt{1-M_{11}M_{22}}e^{-i\phi} & M_{22}
\end{bmatrix}
\end{split}
\end{equation}
and 
\begin{equation}
M_{11}M_{22}\leq 1.
\end{equation}
 Here $\phi$ corresponds to the phase of  right reflection coefficient   $r_R$, with $\phi\in[0,2\pi]$.
Further, the transmission coefficient   $t$, left reflection coefficient $r_L$, and right  reflection coefficient $r_R$ 
can be expressed in terms of parametrization: $t=\frac{1}{M_{22}}$, $r_L=\frac{\sqrt{1-M_{11}M_{22}}}{M_{22}}e^{-i\phi}$, and $ r_R=\frac{\sqrt{1-M_{11}M_{22}}}{M_{22}}e^{i\phi}$.
By one observation, we can find that 
$t=t^{*}$ and $r_L=r_R^{*}$ are general properties for any APT-symmetric systems, already stated in \cite{APT1}.
The former relation indicates that the transmitted wave with respect to the incident wave has either in phase or out of phase, while the latter relation reveals the left and right reflectances defined by $R_L=\vert r_L\vert^2$ and $R_R=\vert r_R\vert^2$ are same in magnitudes but their phase  relation requires $Arg[r_L]=-Arg[r_R]$.

 \begin{figure}[t]
\centering
\includegraphics[width=0.4\textwidth]{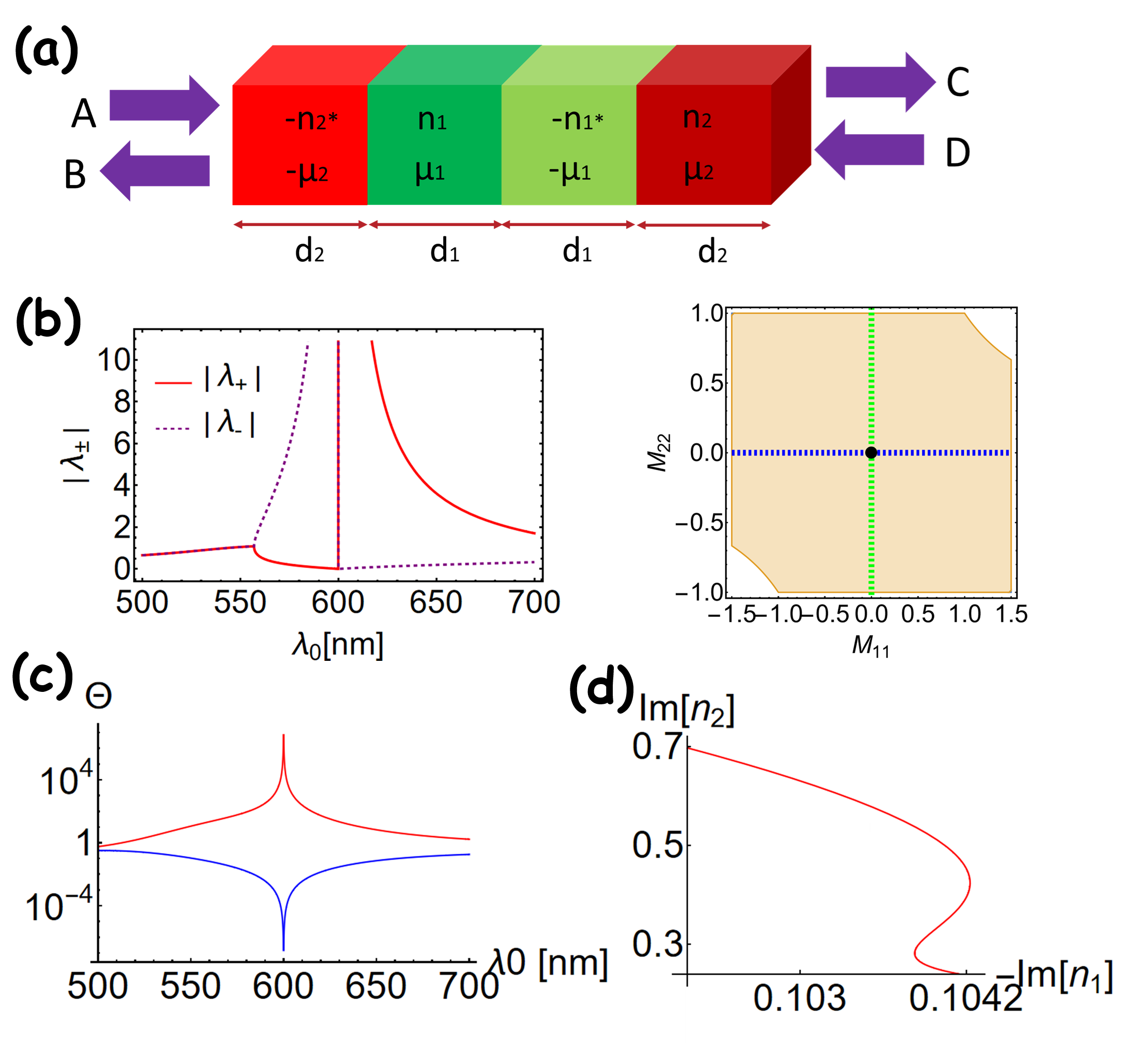}
 \caption{(a) Schematic of a four-slab heterostructure. Magnitudes of the associated scattering eigenvalues $\vert \lambda_{\pm}\vert$ are calculated in left of (b). With the parametric space, we analyze the four-slab heterostructure marked by a black dot in right of (b). (c) $\Theta$ with respect to $\lambda$. Here the blue curve  is obtained by a coherent relation for two incident waves with $D=AM_{21}$, while the red one is from one single port excitation. With fixed $Re[n_1]=-2.1$, $Re[n_2]=-1.5$, $d_1=500 [nm]$, and $\lambda_0=600 [nm]$, we consider the imaginary part of $n_1$ and $n_2$ for CPAL by tuning $d_2\in[500 nm, 550 nm]$ as shown in (d).}
 \end{figure}
 
\textit{Symmetry phase, Exceptional point, and  symmetry broken phase-}
The scattering matrix $S$ is described by $[B,C]^T=S[A,D]^T$, where 
\begin{equation}
\begin{split}
S=\begin{bmatrix}
r_L & t\\
t & r_R
\end{bmatrix}=\begin{bmatrix}
\frac{\sqrt{1-M_{11}M_{22}}}{M_{22}}e^{-i\phi} & \frac{1}{M_{22}}\\
\frac{1}{M_{22}} &\frac{\sqrt{1-M_{11}M_{22}}}{M_{22}}e^{i\phi}
\end{bmatrix}
\end{split}
\end{equation}
is a symmetric matrix and can be further expressed in terms of parametrization.

 The corresponding scattering eigenvalues $\lambda_{\pm}$ and eigenvectors $\vert \lambda_{\pm}>$ are
$\lambda_{\pm}=Re[r_R]\pm\sqrt{t^2-Im^2[r_R]}$ and $\vert \lambda_{\pm}>=[-t,iIm[r_L]\mp\sqrt{t^2-Im^2[r_R]}]^T$, respectively.
With the benefit of parametrization, it provides an alternative way to see $S$ having pseudo-Hermitian property, ensuring an existence of  a spectra for real eigenvalues \cite{pseudo}.
With analysis of the square root of $t^2-Im^2[r_R]$, we observe that APT symmetric systems supports two scattering phases: symmetry phase and symmetry broken phase.
For the former,  it requires $t^2>Im^2[r_R]$ which is equivalent to $(\cos^2\phi+\sin^2\phi M_{11}M_{22})/M_{22}^2>0$ by parametrization, resulting in two real and distinct $\lambda_{\pm}$.
For the latter, it requires $t^2<Im^2[r_R]$ which is equivalent to $(\cos^2\phi+\sin^2\phi M_{11}M_{22})/M_{22}^2<0$ by parametrization, resulting in $\lambda_{+}=\lambda_{-}^{*}$.
In a spontaneous phase transition, it corresponds to an EP with two scattering eigenvalues and scattering eigenvectors being coalesced.
Here it corresponds to $(\cos^2\phi+\sin^2\phi M_{11}M_{22})/M_{22}^2=0$ by parametrization.

Based on these outcomes and Eq.(2), we propose a parametric space in terms of $M_{11}$, $M_{22}$, and $\phi$ as shown in Fig. 1 (b).
Here the regions in blue color denote  symmetry phase, while the region in yellow color denote symmetry broken phase.
In between, it is EP, marked by purple lines as shown in Fig. 1 (c).
To explore the scattering phases in a more detail, we plot the 2D parametric spaces in terms of $M_{11}$ and $M_{22}$ with respect to a variety of $\phi$ as shown in Fig. 1 (c), where $\phi$ involves the scattering phase. 
With the parametric space, we further plot the transmittance $T=\vert t\vert^2$ and reflectance $R_R=R_L=R$ in Fig. 1 (d).
Obviously, an experimental measurement for T or R by a single port excitation can not determine their scattering phases, unlike PT-symmetric circumstances \cite{yidong,Lee,PT3}.

\textit{Scattering properties-}
APT-symmetric systems enables total transmission, lasing occurred at symmetry phase, CPA occurred at symmetry phase, that already got well stated in Refs.\cite{APT1,APT2}.
Alternatively, with the parametric space, we can uncover undiscovered wave scattering phenomena.

In absence of gain and loss, APT-symmetric systems satisfy the unitary relation, i.e., $S^{\dagger}S=I$. 
It leads to $r_R=r_L=0$  and $t=1$.
With complementary media embedded, the systems exhibit bidirectional total transmission (BTT)\cite{APT1,APT2,complementary}.
By parametrization, it corresponds to $M_{11}=1$ and $M_{22}=1$. 
We mark it by a red point in Fig. 1 (c).
Next, in lasing, the system can generate outgoing coherent waves without any incident waves.
The system needs gain materials embedded.
It  corresponds to $M_{22}=0$ and, at least, a pole occurred in one of scattering eigenvalues $ \lambda_{\pm}$.
We mark $M_{22}=0$ by a black dashed line in Fig.1 (c).
Moreover, there retains a coherent relation for associated outgoing waves by $C=M_{12}B=e^{i\phi}B$.
Next, we observe that two EP lines depicted by purple color in the parametric space can involve the black dashed line for $M_{22}=0$ when $\phi=\pm\frac{\pi}{2}$, as indicated in the bottom of Fig. 1 (c).
To prove this finding, we use parametrization to consider the criterion of $t^2-Im^2[r_R]$ and apply
 L'Hopital's rule:
 \begin{equation}
 \lim_{M_{22}\rightarrow 0}t^2-Im^2[r_R]\vert_{\phi=\pm\frac{\pi}{2}}= \lim_{M_{22}\rightarrow 0}\frac{M_{11}M_{22}}{M_{22}^2}=0,
 \end{equation}
satisfying the EP condition.
Therefore, lasing can occur not only at symmetry phase but also at EP.
For the latter, there are two poles simultaneously occurred in scattering eigenvalues $\lambda_{\pm}$.
To support this result,
we design a simple  bi-slab heterostructure  as shown in left of Fig. 3 (a), with gain embedded.
We exploit the following  parameters by relative permeability $\mu_{1}=-2.16$, refractive index $n_{1}=-2.12-0.3i$, slab length $d=447.66 [nm]$, and operating wavelength of free space $\lambda=610 [nm]$.
We calculate T, $R_L$, and $R_R$ with respect to $d$ in Fig. 2 (a), where we can observe two large tips at $\lambda=450 [nm]$, supporting the formation of lasing.
Any emergence of a pole in scattering matrix S would accompany by a pole in $T$ and $R$ spectra.
Then, we estimate $\vert \lambda_{\pm}\vert$ in the left of Fig. 2 (a). 
We can see that as $d<450 nm$, the system is in symmetry phase, while as $d>450 nm$, the system is in symmetry broken phase.
With $d=450nm$, the system is in EP.
Alternatively, we analyze the system by the parametric space as a black dot of Fig. 2 (b), supporting lasing in EP.

Waves propagating in NIMs form anti-parallel direction between wave vector and Poynting vectors.
Due to this event, in APT-symmetric systems, the phase accumulated from propagating path length can be null, leading to $t=t^{*}$. 
On the other hand, as APT-symmetric systems is served as a laser oscillator, this property results in an assignment of mode order lost.
We first consider a conventional optical feedback system, where a gain is surrounded by two identical lossless mirrors, as shown in the left of Fig. 2 (c).
We adopt multiple scattering formalism to analyze our case, which we regard a resultant standing wave in a cavity  as a sum of travelling waves \cite{book1}.
Therefore, after one-round trip, a travelling wave $U_m$ becomes $U_{m+1}$ with a relation of
$U_{m+1}/U_m=e^{i2n_g(\omega)k_0L}r_M^2$.
Here we denotes $m=1,2,...$  as an integer index  for self-reproducing waves, $n_g(\omega)$ is complex refractive index of gain, $r_M$ is reflection coefficient of a mirror, $k_0$ is wavenumber of free space, and $L$ is  length of a cavity.
To reach a lasing threshold, there would satisfy phase and magnitude conditions by $2k_0LRe[n_g]+2Arg[r_M]=2\pi m$ and $\vert r_M\vert^2e^{-2k_0LIm[n_{g}]}=1$, respective.
The discrete order $m$ in the phase condition is closely related to the longitudinal resonant frequencies.
By same analysis approach, we consider  a bi-slab heterostructure having APT-symmetry surrounded by same mirrors, shown in Fig. 2 (c).
We formulate the necessary condition for self-reproducing scattering waves to reach a lasing threshold
\begin{equation}
e^{2ink_0\frac{L}{2}}t_1e^{-2in^{*}k_0\frac{L}{2}}t_2r_1r_2=1
\end{equation}
here $L$ is total cavity length, $t_1$ ($t_2$) is complex transmission from left to right (right to left) in the interface of slabs, $r_1$ ($r_2$) are complex reflection coefficients from the left (right) mirrors \cite{note1}.
Now, in the phase condition, we state $Arg[t_1]+Arg[t_2]+Arg[r_1]+Arg[r_2]=2\pi m$. 
Here $m$ denotes an integer number.
Obviously, the parameter of the optical path length is lost, 
 resulting in  an assignment of order index lost.
With fixed real part of refractive index $Re[n]$, operating wavelength of free space $\lambda_0$, and relative permeability $\mu$, we calculate  a lasing threshold of $Im[n]$ with respect to a slab length $d$ in the left of Fig. 2 (f).
Unlike conventional discrete resonant points in spectra, APT-symmetric systems exhibit a continuous spectra.

In CPA, appropriate input coherent waves interacted with a cavity system would be totally absorbed, resulting in an absence of outgoing waves, i.e.,  $B=0$ and $C=0$.
The CPA condition is $M_{11}=0$ in the transfer matrix, while at least, it has one zero of scattering eigenvalues $\lambda_{\pm}$.
CPA is time reversal counterpart of lasing, while
its composition needs lossy materials \cite{CPA}.
A coherent relation for two input waves is necessary to exhibit CPA phenomenon, i.e., $D=AM_{21}$.
Again, with the benefit of parametric space, we observe that  CPA can occur at symmetry phase and EP.
To prove the latter one, we use parametrization to consider the criterion of $t^2-Im^2[r_R]$ and apply
 L'Hopital's rule:
 \begin{equation}
 \lim_{M_{11}\rightarrow 0}t^2-Im^2[r_R]\vert_{\phi=\pm\frac{\pi}{2}}= \lim_{M_{11}\rightarrow 0}\frac{M_{11}M_{22}}{M_{22}^2}=0,
 \end{equation}
satisfying the EP condition.
As a result, CPA can occur not only at symmetry phase but also at EP.
We consider a bi-slab heterostructure with loss alone.
The parameters we use are $\mu_{1}=-2$, $n_{1}=-1.91+0.33i$, $d=313.85[nm]$, and $\lambda=530[nm]$.
We calculate $\vert \lambda_{\pm}\vert$ with respect to d in Fig. 2 (d).
We observe that there have two dips simultaneously occurred at $d=320 [nm]$, corresponding to two zeros of scattering eigenvalues.
When $d<320 nm$, the system is at symmetry phase, while when $d>320 nm$, the system is at symmetry broken phase.
We analyze the system at $d=320 [nm]$ with the parametric space, marked by a black dot as shown in Fig. 2 (e), supporting CPA at EP.
Due to balanced PIMs and NIMs, there has expected to see a continuous spectra of $Im[n]$ with $d$ with fixed $Re[n]$, as shown in the right of Fig. 2 (f). 

APT symmetric system enables a simultaneous CPA and lasing (CPAL), as clearly indicated in the parametric space of Fig. 1 (c).
Interestingly, the CPAL of APT symmetric systems has not been studied yet.
To have CPAL, it requires $M_{11}=0$ and $M_{22}=0$, which have one zero and one pole simultaneously occurred in $\lambda_{\pm}$. 
We design such system by a four-slab heterostructure shown in right of Fig. 3 (a), with one group of materials being gain and another being loss.
The parameters we use are $\mu_1=-1.3$, $n_1=-0.591-0.1i$, $l_1=570 nm$, $\mu_2=-1.36$, $n_2=-0.598+0.5i$, and $l_2=400 nm$.
We calculate $\vert\lambda_{\pm}\vert$ with respect to an operating wavelength $\lambda_0$ within $[500 nm , 700 nm]$. 
We observe that when $\lambda_0<550 nm$ and $550 nm<\lambda_0<700$, the system is in broken symmetry phase  and symmetry phase, respectively.
When $\lambda_0=550 nm$, the system is in EP.
As $\lambda_0=600 nm$, one scattering eigenvalue is zero and another is pole. 
At $\lambda_0=600 nm$, we analyze the system by the parametric space marked by a black dot in right of Fig. 3 (b), where this system exhibits CPAL as as well as is in symmetry phase.
To observe CPA phenomenon, there needs to import proper incident waves with a coherent relation of $D=AM_{21}$.
Any incident waves deviated from this relation would have been amplified.  
Following the approach in Ref.\cite{CPAL}, we define a non-dimensional ratio by total out-going intensities over total incident ones, i.e., $\Theta=\frac{\vert B\vert^2+\vert C\vert^2}{\vert A\vert^2+\vert D\vert^2}$.
At a single port excitation, i.e., $A=1$, we have $\Theta=T+R$.
We calculate $\Theta$  with respect to $\lambda_0$ in Fig. 3 (c), where there has a tip at $\lambda_0=600 [nm]$, corresponding to light amplification.
However, for two incident waves with a coherent relation $D=AM_{21}$, we observe a dip at $\lambda_0=600 [nm]$, corresponding to light totally attenuation.
Unlike  PT-symmetric systems where one zero and one pole in scattering eigenvalues appear in a pair, APT-symmetric systems can enable CPA alone or   lasing alone.
Due to balanced PIMs and NIMs for APT-symmetric systems, we can find a continuous spectra for $Im[n_{1}]$ and $Im[n_{2}]$ with respect to $d_2$ with fixed $Re[n_1]$ and $Re[n_2]$, as shown in Fig. 3 (d).
In a perspective of electromagnetic field energy, the permeability must be dispersive in operating spectra, leading to unavoidably material loss.
However, it can be overcome by gain materials embedded, such as Rhodamine 800 dye \cite{gain1,gain2}.

In conclusion, with the APT-symmetry property and optical reciprocity, we propose the parametric space to indicate not only symmetry phase, EP, and symmetry broken phase, but also transmittance and reflectance.
By  the parametric space, we observe that APT-symmetric systems can enable not only CPA and Lasing at EP, but also CPAL.
We design associated heterostructures to verify our findings. 
Attributed from balanced NIM and PIM, the phase accumulated from optical propagating path is null, resulting in a null transmission phase and an assignment of mode order lost.



\begin{thebibliography}{1}
\bibitem{note1}
Between the slab interface, we formulate $r_1$ and $r_2$ by $-\frac{iIm[n]}{Re[n]}$ and $\frac{iIm[n]}{Re[n]}$, respectively. 
We  assume $Re[n]>>Im[n]$, so that $\vert r_1\vert <<1$ and $\vert r_2\vert <<1$.
Therefore, we ignore reflected waves in left and right  raised by the interface.
\end{thebibliography}
\end{document}